\documentclass[sn-mathphys]{sn-jnl}

\jyear{2021}%

\theoremstyle{thmstyleone}%
\theoremstyle{thmstyletwo}%

\theoremstyle{thmstylethree}%

\begin{document}

\title[Enhance. laser-induced excit. prob.]{Enhancement of the 
laser-induced excitation probability of the hyperfine 
ground state of muonic hydrogen by a multipass cavity setup}


\author*[1,2]{\fnm{Rakesh Mohan} \sur{Das}}\email{rakesh.dasfpy@kiit.ac.in}

\author[3]{\fnm{Masahiko} \sur{Iwasaki}}


\affil*[1]{\orgdiv{School of Applied Sciences}, \orgname{KIIT University}, \orgaddress{\street{Patia}, \city{Bhubaneswar}, \postcode{751024}, \state{Odisha}, \country{India}}}

\affil[2]{\orgdiv{Center of Excellence in High Energy and 
Condensed Matter Physics, Department of Physics}, \orgname{Utkal University}, \orgaddress{\street{Vani Vihar}, \city{Bhubaneswar}, \postcode{751004}, \state{Odisha}, \country{India}}}

\affil[3]{\orgdiv{The Institute of Physical and Chemical Research}, \orgname{RIKEN}, \orgaddress{\street{2-1 Hirosawa, Wako}, \city{Saitama}, \postcode{351-0198}, \state{}, \country{Japan}}}


\abstract{We study the enhancement of the magnetic dipole induced 
excitation probability of the hyperfine ground state of 
Doppler-broadened muonic hydrogen ($p \mu^{-}$)
by a nanosecond laser pulse in the mid-infrared range with Gaussian 
temporal shape such 
that the pulse bandwidth is broader than the Doppler width at 
10~K. The enhancement is achieved by shrinking the 
cross-section of the laser pulse and placing the muonic hydrogen 
medium in a  multipass cavity, while preserving the total 
irradiated target volume. 
We numerically solve a set of Maxwell-Schr$\ddot{\rm o}$dinger 
equations to obtain the excitation probability and the total 
efficiency for various densities of the muonic hydrogen atomic 
medium and at various positions in the multipass cavity. For the 
typical range of densities of muonic hydrogen atoms 
at major proton accelerator facilities such as the J-PARC 
(density $\sim$ $10^5$cm$^{-3}$), 
the laser propagation effect is insignificant. For such cases, 
the total efficiency increases by an order of two for 100 
reflections with a uniform polarization. 
If the density exceeds the 
value of $10^{17}$cm$^{-3}$ as might be in the future advances, 
the laser propagation effect has to 
be taken into account, and the total efficiency decreases with 
the number of reflections giving rise to a pulsed polarization of 
the beam. Our study can serve as a guideline for the development of 
a polarized muonic beam for a precise measurement of 
the ground state hyperfine splitting of muonic hydrogen, or  for 
$\mu$SR experiments.}




\maketitle

Amongst all the exotic atoms, muon ($\mu^{\pm}$) prevails 
as the probing tool 
as far as the theoretical studies and experiments in atomic and 
nuclear physics are concerned. This is due to its relatively 
easier availability and longer lifetime of 2.2~$\mu$s. 
Charged pion which is often the next in line amongst the exotic atoms, 
has a lifetime of 26~ns, making it very short-lived. Besides, 
unlike the pion, 
the muon is not influenced by the strong force of the atomic nuclei 
in the lattice, and hence has higher penetration depth.
Muon has been studied extensively due to its relevance 
in the Standard model of elementary particles, and its usefulness in 
various applications such as $\mu$SR experiment, and 
muon-catalyzed fusion, etc \cite{Gorringe2015, PohlNature2010, 
AntogniniSci2013, Bakule2004, 
Miller2007, Lancaster2007, Bungau2014, Nagamiya1975, Reotier1997, 
Balin2011}.
Commonly, for the application purpose of muon (whether 
positively or negatively charged), its kinetic energy is reduced 
from as high as a few MeV after being ejected from the particle 
accelerator, down to a few keV $\sim$ a few eV. 
In this process, positive muons capture electrons to form muonium 
atoms while negative muons are captured by protons to form muonic 
hydrogen atoms. In both the cases of the recombination processes, 
the degree of polarization, which is 100 \% upon birth, is 
significantly lost which needs to be spin-repolarized.

\begin{figure*}[t]
\centerline{
\includegraphics[width=0.9\textwidth]{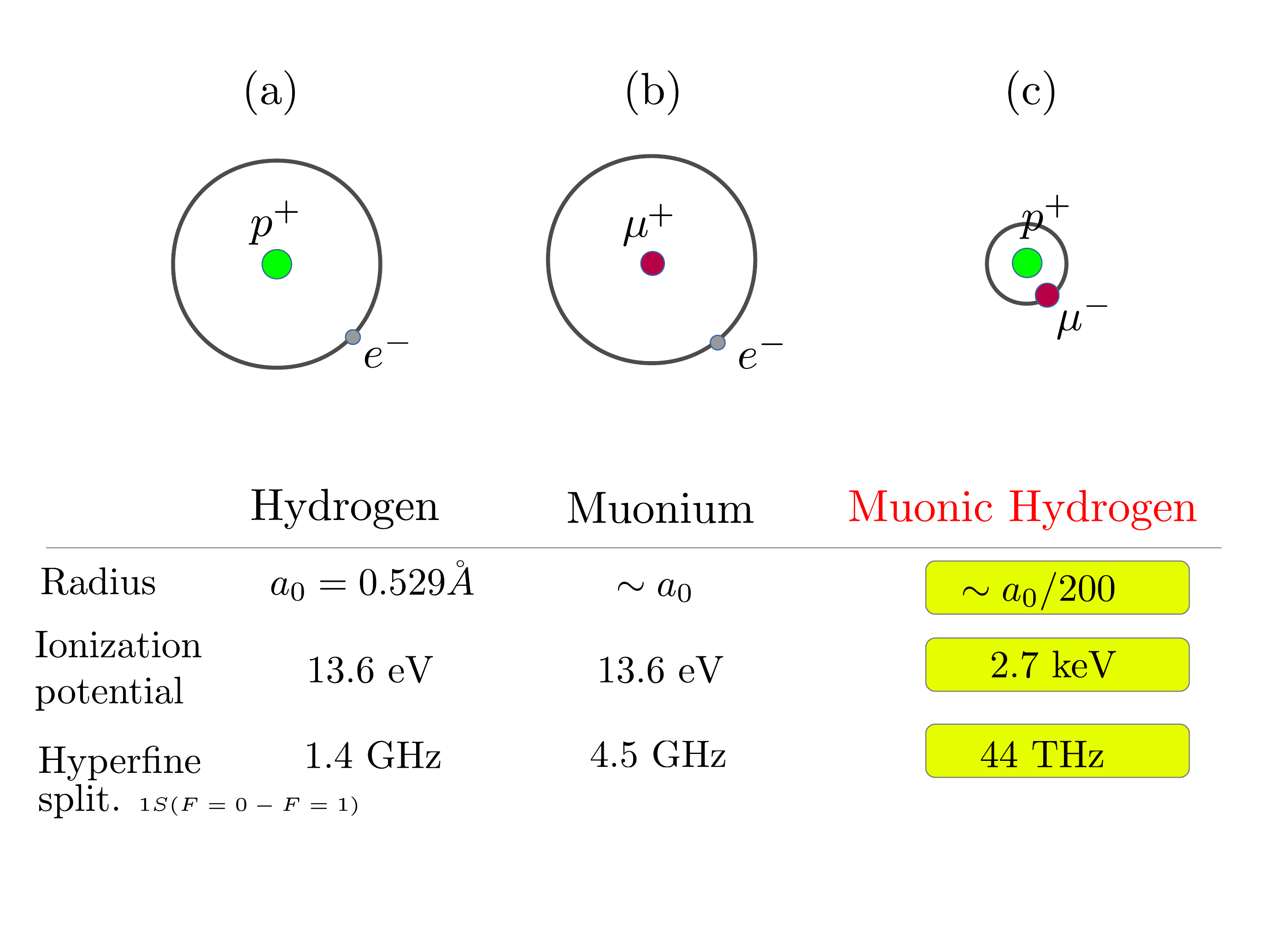}
}
\caption{(Color online) Comparison of the hydrogen and 
hydrogen-like systems involving 
positive and negative muons. }
\label{fig1}
\end{figure*}

Several works have been done on obtaining slow positive muon beam 
\cite{Nagamine1995, Morenzoni1994, Bakule2003, RT-muonium2014,
NakajimaOptExp2010, Nakajima2012,DasJOSAB2015,DasJOSAB2018}.
We turn our focus on the muonic hydrogen which 
has proved to be a suitable 
candidate to precisely determine the size of proton. 
This is because of the proximity of the negative muon to the proton 
that amplifies the atomic properties of the 
muonic hydrogen due to the presence of the negative muon 
(200 times heavier particle than the electron) in the 
atomic orbit with a radius of about 1/200~th of 
that of hydrogen and muonium. 
For comparison, we have used the figure from our previous work 
\cite{DasOptik2020} which summarizes the various atomic parameters 
of hydrogen, muonium, and muonic hydrogen in Fig.~\ref{fig1}.

The famous study on the Lamb shift 
($2S_{1/2} - 2P_{1/2}$ energy difference) of muonic hydrogen 
at PSI \cite{PohlNature2010} reported that the proton radius is 
about $4 \%$ smaller than the widely accepted value. This puzzle 
has spearheaded many works on spectroscopy and particle physics 
\cite{YamamotoNIMS2021, SERAFINI2018309, BNLmuoncollider2019}. 
Along this line, Adamczak \textit{et al.} proposed a 
method of determining the proton size with accuracy by an 
independent measurement of the hyperfine splitting 
($1S (F=0) - 1S (F=1)$) of the ground state 
of muonic hydrogen \cite{AdamczakNIM2012, 
Sapirstein1990, Eides2001}. The $1S (F=0) - 1S (F=1)$ transition  
wavelength is in the mid-infrared range (6.76~$\mu$m), is 
dipole-forbidden, and the process involves magnetic dipole 
transition which is several orders of magnitude weaker
than the electric dipole transition.
The experimental scheme was originally proposed by Bakalov 
{\it et al.} \cite{Bakalov1993}, in which 
negative muons are slowed down and stopped in a pure hydrogen target 
between two parallel gold or 
aluminium plates to form muonic hydrogen. Recently,  
we also have studied the laser-induced 
hyperfine ground state excitation probability 
of Doppler-broadened muonic hydrogen for various pulse parameters 
that can serve as a guideline for the development of the relevant 
laser source, depending on whether the purpose of 
exciting muonic hydrogen is measuring the ground state hyperfine 
splitting of muonic hydrogen with precision, or 
producing spin-polarized muonic hydrogen 
beam \cite{DasOptik2020}. 
Adamczak {\it et al.} have 
proposed a scheme of placing the target inside a 
multipass cavity in order to 
increase the excitation probability which is otherwise merely 
$1.2 \times 10^{-5}$ (too low for the proposed experiment 
to be feasible) with a laser pulse of long duration at 
6.76~$\mu$m with 0.25~mJ pulse energy in a single interaction 
of the laser pulse with the muonic hydrogen beam at a temperature 
of 300 K \cite{AdamczakNIM2012}.
Undoubtedly, this experiment comes with huge technical 
challenges. 
From a theoretical point of view, a study of the elementary 
challenges such as the deterioration 
of the laser field profile as it propagates through the muonic 
hydrogen medium inside the multipass cavity for various ranges 
of medium densities, is essential for completeness. For 
the muon beam facilities around the world at present, the 
achievable beam density is not high enough for any laser 
propagation effect to arise. For example, at the J-PARC, 
the density of the muon beam is typically $10^5$cm$^{-3}$.  
Nevertheless, it is important 
to study the laser propagation effect on the polarization of the 
muons, as the density is increased, as might be possible 
by the technological development in future. 
Besides, the prospects and limitations in tailoring the 
temporal profile of the muonic hydrogen beam in such a 
multipass cavity setup, needs to be discussed. 
We note that, pulsed muon and CW muon sources that have been 
developed at various facilities across the world, have their 
own advantages and disadvantages depending upon the purpose  
of the muon source. For example, the PSI and 
TRIUMF \cite{ForoughiHyper2001, WalterHighInten2001} are both home 
to intense continuous muon beam 
which is suitable for spectroscopy experiments, whereas 
KEK, RIKEN-RAL and J-PARC MUSE \cite{MatsuzakiNIMS2001, 
jparcmuse2012} serve as the sources of pulsed 
muon beams which are better suited for $\mu$SR experiments. 
It is to be seen in this context, 
how the multipass cavity setup might  
play its role in shaping the temporal profile of the muonic hydrogen 
beam.

In the present work, we study the excitation probability of 
the muonic hydrogen by nanosecond laser pulses in a multipass 
cavity setup and investigate the conditions under which 
the excitation probability is enhanced which leads to increased 
polarization of muons. The bandwidth of the laser pulse we 
consider in our study is broader as compared to the Doppler width 
of the atoms. The results we present in this work would be useful  
in the development of experimental setup for polarized muonic beam.

\section{System Description} 
\label{system}

\begin{figure*}[!ht]
\centerline{
\includegraphics[width=0.6\textwidth]{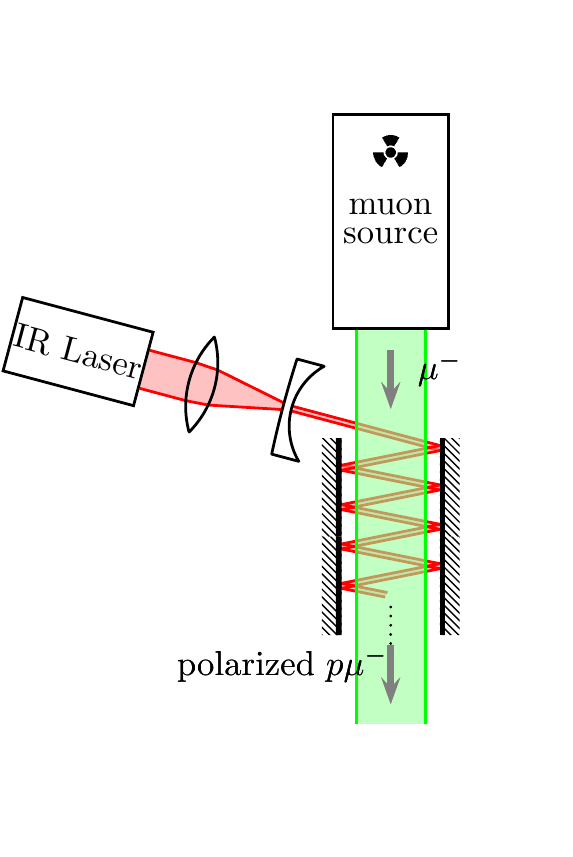}
}
\caption{
(Color online) 
Schematic diagram showing the muonic hydrogen atoms inside a 
multipass cavity with mirrors of perfect reflectivity. An infrared 
laser pulse is shrunk in its beam cross-section and goes 
through a number of reflections such that the interaction volume of 
the muonic hydrogen and the laser pulse is preserved.
}
\label{fig2}
\end{figure*}

Figure~\ref{fig2} shows the schematic diagram of the 
system we have considered. 
We consider the hyperfine sublevels of the ground ($1s$) 
state of the 
muonic hydrogen. A laser beam in the mid-infrared range (6.7~$\mu$m) 
is shrunk by reducing its cross section by a factor $k$ (thereby 
increasing the intensity by the same factor), and projected into a 
multipass cavity with perfect reflectivity. The muonic hydrogen beam 
is passed through the multipass cavity providing $k$ reflections 
of the 
laser beam, thereby preserving the total irradiated target volume.  
The assumption of a 
multipass cavity with perfect reflectivity is for the sake of 
simplicity given that the intensity of the laser beam is barely 
attenuated by the mirrors of reflectivity as high as 99.999~\% for 
the number of reflections considered in this work.
In steady state, the slow beam of muonic hydrogen, after the 
fast muons being captured by the hydrogen atoms inside the cavity, 
can be treated to be frozen in space when compared to the laser 
pulse passing through the multipass cavity.

We assume a Gaussian pulse which 
excites the muonic hydrogen from the 
hyperfine singlet state, $1s (F=0)$, to the triplet state, 
$1s (F=1)$, 
through magnetic dipole ($M1$) interaction. We note that, the 
hyperfine transition is electric dipole forbidden and it is the 
magnetic dipole component which has the most significant contribution 
in the laser-matter interaction.  
The magnetic field of the pulse with central frequency $\omega$ is 
written as

\begin{align}
\label{eqn:field1}
\mathcal{B}(t) & =  B(t) \exp{\left[-i\omega t \right]} + \mathrm{c.c.},
\end{align}

\noindent 
where $B(t)$ is the magnetic field amplitude of the pulse. The 
magnetic transition dipole moment of the muonic hydrogen atom, 
$M$, calculated by us in our previous work is 
$\frac{1}{\sqrt{6}} \frac{e\hbar}{2m_\mu}$ \cite{DasOptik2020}. 
The Rabi frequency $\Omega$, corresponding to the magnetic dipole transition in S.I. unit, is  

\begin{align}
\label{eqn:rabins}
\Omega ({\rm ns}^{-1}) & = 1.77 \times 10^{-6}\sqrt{I(t)}
\end{align}

\noindent 
where $I(t)$ is the laser intensity in units of W/cm$^2$.
We employ a set of Maxwell-Schr$\ddot{\rm o}$dinger equation, 
which can be easily obtained by 
coupling the probability amplitudes, $a_0$ and $a_1$, of the 
hyperfine states, $1s (F=0)$ and $1s (F=1)$, respectively and 
Maxwell's equations, which reads

\begin{align}
\label{eqn:amp1}
\frac{\partial}{\partial t} a_0(z,t) = i \Omega^{*}(z,t) a_1(z,t) ,
\end{align}

\begin{align}
\label{eqn:amp2}
\frac{\partial}{\partial t} a_1(z,t) = i \Omega(z,t) a_0(z,t) ,
\end{align}

\begin{align}
\label{eqn:amp3}
\frac{\partial}{\partial z} \Omega(z,t) =  -2i\alpha a_1^{*}
(z,t)a_0(z,t).
\end{align}

\noindent 
where $\alpha = N \omega |M|^2 \mu_0 c /2$, is defined as the laser 
propagation coefficient of the muonic hydrogen medium, $N$ 
being the density of the medium \cite{Buica2014}. While obtaining 
the above equations, we have neglected the spontaneous decay 
rate of the muonic hydrogen for the range of time duration considered.
Although the slow muonic hydrogen atoms formed are unpolarized, 
i.e. 50$\%$ of the atoms are in the state $1s(F=0)$ and 50$\%$ are in 
state $1s(F=1)$, the initial condition that all the atoms are in the 
singlet state, $1s(F=0)$, simplifies the study without the 
loss of generality. We numerically calculate the Doppler-averaged 
spin-flip probability, $P$, of the ground hyperfine state, from 
$1S (F=0)$ to $1S (F=1)$, of the muonic hydrogen, 
taking into account the laser 
propagation effect of the medium on the laser pulse as it 
passes through the multipass cavity. 

\begin{figure*}[!ht]
\centerline{
\includegraphics[width=0.9\textwidth]{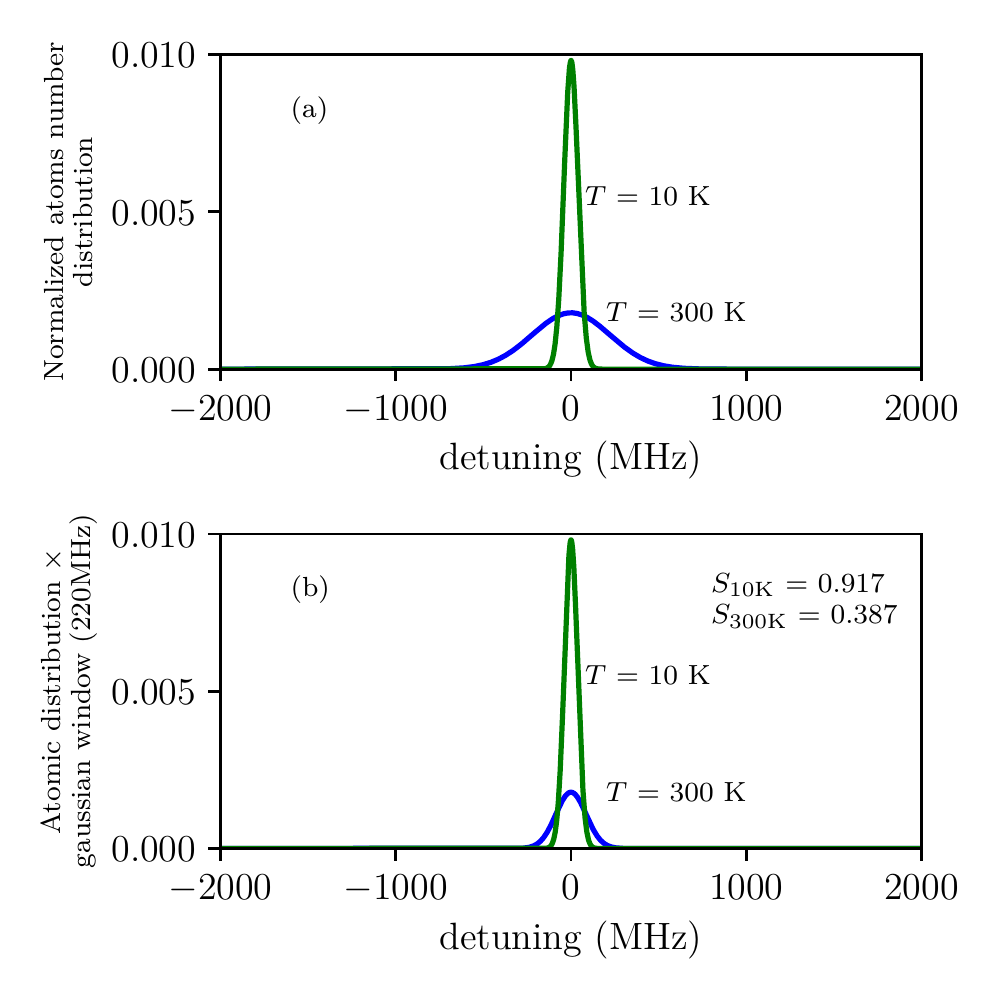}
}
\caption{
(Color online)
(a) Normalized distribution of the number of Doppler-broadened 
muonic hydrogen atoms at a temperature of 300~K (blue line) and 
10~K (green line). The area under each of the lines is unity. 
(b) Same as panel (a) multiplied with a 
Gaussian window function of width 220~MHz, which corresponds to  
the spectral width of a 2~ns Gaussian pulse. The ratio of the 
areas under the green and red line is 0.917/0.387 = 2.3. 
}
\label{fig3}
\end{figure*}

\begin{figure*}[!ht]
\centerline{
\includegraphics[width=0.7\textwidth]{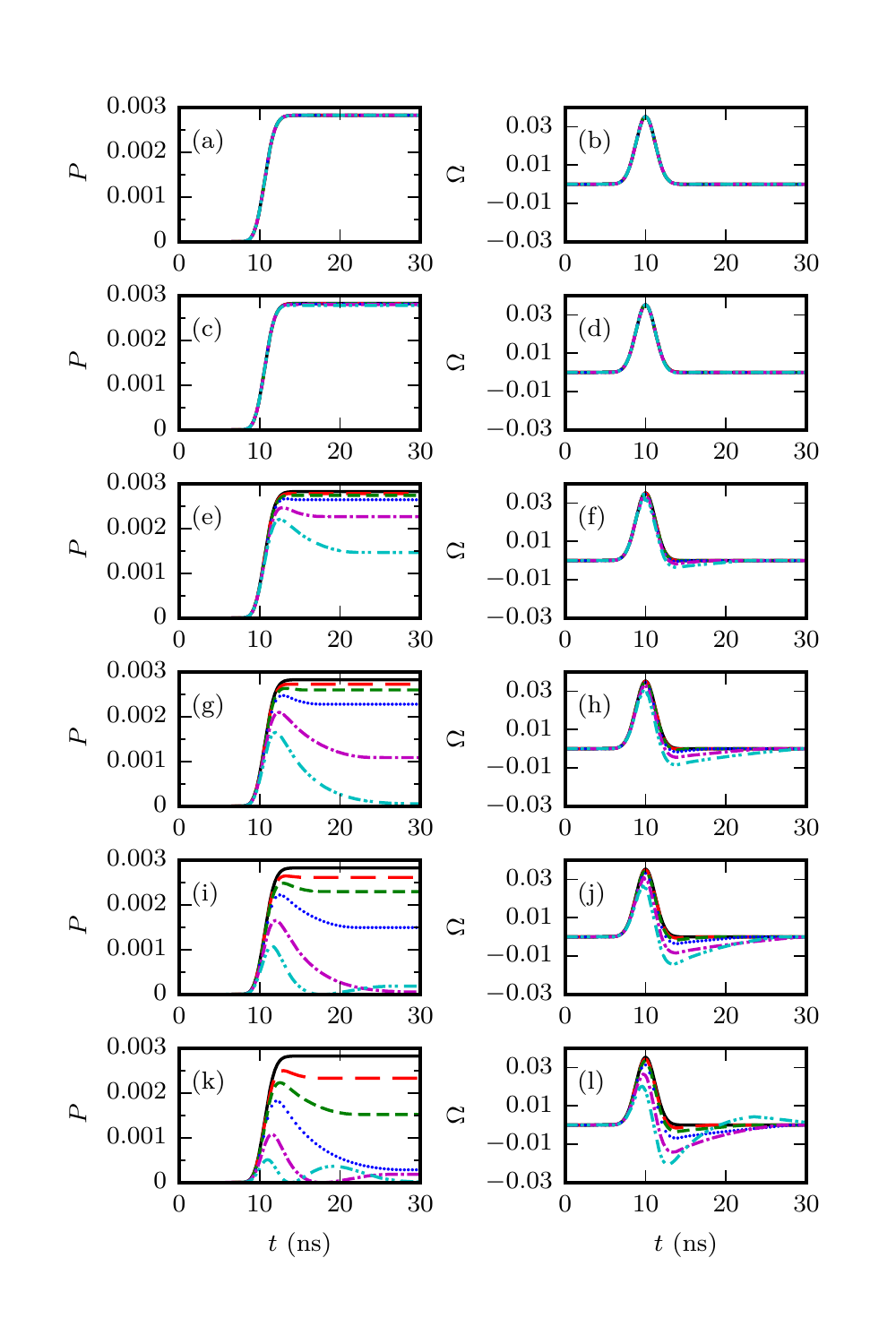}
}
\caption{
(Color online) 
Temporal variation of the population of the ground hyperfine state, 
$1s(F = 1)$ and the laser field, $\Omega$ with the medium propagation coefficient, 
$\alpha = 4 \times 10^{-18}$ mm$^{-1}$ [(a), (b)], 
$4 \times 10^{-6}$ mm$^{-1}$ [(c), (d)], 
$4 \times 10^{-5}$ mm$^{-1}$ [(e), (f)],
$1 \times 10^{-4}$ mm$^{-1}$ [(g), (h)],
$2 \times 10^{-4}$ mm$^{-1}$ [(i), (j)], and
$4 \times 10^{-4}$ mm$^{-1}$ [(k), (l)], at different reflection 
numbers, $k = 1$ (black solid line), 5 (red long-dashed line), 
10 (green short-dashed line), 20 (blue dotted line), 
50 (magenta dot-dashed line), and 100 (cyan double dot-dashed 
 line). 
The laser pulse intensity is $1 \times 10^6$ W/cm$^{2}$ and the 
pulse duration is 2~ns for all the graphs.
}
\label{fig4}
\end{figure*}

\section{Results and Discussions} 
\label{result}
At the outset, it should be noted that we confine our study to a 
peak intensity, $I_0 = 1 \times 10^6$ W/cm$^2$ and a duration, 
$\tau = 2$ ns of the laser pulse.
At resonance, the Doppler-averaged spin-flip probability of 
the muonic hydrogen atoms at 10~K, is enhanced by a factor 
of 2.3 as compared to that at 300~K. The lowest temperature 
the system can go is 10~K as below which molecular $p \mu p$ 
are formed. Hence, we also confine 
our study to the muonic hydrogen medium at a temperature of 
10~K.
Additionally, at 10~K, the Doppler 
broadening of the medium is about 100~MHz which is less than 
half of the 
bandwidth of a 2~ns pulse, namely  
220~MHz. As the bandwidth of the pulse covers the Doppler width 
of the medium fairly enough, we can safely calculate 
the spin-flip probability at resonance without taking the 
Doppler average and 
apply a Gaussian window function with the parameters 
equal to that of the laser pulse to effectively calculate the 
Doppler-averaged spin-flip probability, $P$ at 10~K, which saves 
computation time, without loss of accuracy. In Fig.~\ref{fig3}~(a), 
we show the normalized Doppler distribution of muonic hydrogen 
atoms at 300~K and 10~K, with blue and green solid lines, 
respectively. The area under each of the lines are unity. 
In Fig.~\ref{fig3}~(b), we show the normalized 
Doppler distribution multiplied with a Gaussian window function 
of width 220~MHz. The areas, $S_{10 \rm K}$ (= 0.917) and 
$S_{300 \rm K}$ (= 0.387), 
of the modified distribution lines in 
Fig.~\ref{fig3}~(b) represent the number of atoms covered by the 
bandwidth of the laser pulse at temperatures 10~K and 300~K, 
respectively. The ratio of $S_{10 \rm K}$ and $S_{300\rm K}$ 
turns out 
to be 2.3, which is exactly what we have calculated by using the 
amplitude equations of the states and averaging over the 
Doppler distribution of the atoms. This validates our proposed 
method of calculation. Henceforth, in this work, 
$P$ is simply equal to the population of the ground hyperfine 
state, $1s(F=1)$.

We consider a 30~cm long cavity with a width of 10~mm which 
accommodates approximately 100 reflections ($k = 100$) with an 
angular incidence of the laser pulse at an angle of $10$ degrees
as clear from Fig.~\ref{fig2}. It is more 
convenient to use $k$ instead of the otherwise commonly used, 
optical depth, $\alpha z$, where $z$ is the geometrical depth 
that the pulse traverses inside the medium.   
Typical value of the density of the muonic hydrogen atoms, $N$, 
in the cavity, after the slow down process and losses by 
transfer of muons 
to heavier nucleus from the proton, is about 
$10^5$ cm$^{-3}$ which is equivalent to a propagation 
coefficient, $\alpha = 4 \times 10^{-18}$ mm$^{-1}$. At first 
glance, this is extremely 
low to cause any propagation effect on the laser pulse 
for the ranges of optical depths the pulse traverses inside the 
cavity. We calculate the representative values at the midpoint of 
each path between the mirrors, corresponding to $k$. 
In Fig.~\ref{fig4}, we study the time evolution of the 
population of the state $1s(F=1)$, $P$ and temporal variation of the 
laser field, $\Omega$ for increasing values of 
propagation coefficients, $\alpha$ down the columns. 
In each panel, we show the representative values at $k = $1, 5, 
10, 20, 50, and 100 with black solid, red long-dashed, green 
short-dashed, blue dotted, magenta dot-dashed, and cyan double 
dot-dashed lines, respectively. Panels (a), (c), (e), (g), (i), 
and (k) [(b), (d), (f), (h), (j), and (l)] show the 
temporal variation of 
$P$ ($\Omega$) for $\alpha = 4 \times 10^{-18}$, $4 \times 10^{-6}$, 
$4 \times 10^{-5}$, $1 \times 10^{-4}$, $2 \times 10^{-4}$, and 
$4 \times 10^{-4}$ mm$^{-1}$, respectively. Figure~\ref{fig4}~(a) 
shows that $P$ increases by two orders of magnitude 
due to shrinking of the cross-section of the laser pulse. 
It remains same through out the optical path of the pulse as 
the laser field, $\Omega$ is unaffected due to the low muon 
density. 
By observing the Eqs.~\ref{eqn:amp1}-\ref{eqn:amp3}, we note that 
it is the interplay between the pulse area $\sim \Omega \tau$ and 
the optical path, $\alpha z$ 
which dictates the laser propagation effect on the dynamics of the 
system.

The results remain the same as Fig.~\ref{fig4} (a) and (b) for 
the range of values of $\alpha$ until it increases to 
$4 \times 10^{-6}$ mm$^{-1}$ for which temporal variation of 
$P$ and $\Omega$ just barely change with $k$. This is because, 
in that region of $\alpha$, 
$\Omega \tau \gg \alpha z$, hence $\Omega$ effectively remains 
constant. This implies that in this range  of $\alpha$, 
$P$ remains uniform through out the 
cavity as the muonic hydrogen atoms are sparsely populated 
inside the cavity. Therefore, we skip the results for those 
values of 
$\alpha$ and start from $\alpha = 4 \times 10^{-6}$ mm$^{-1}$ 
($N = 10^{17}$~cm$^{-3}$) 
from Fig.~\ref{fig4}~(c) and (d). Undoubtedly, for 
$\alpha = 4 \times 10^{-6}$ mm$^{-1}$, 
$\alpha z$ begins to approach $\Omega \tau$, 
$\alpha z = 0.0004 \Omega \tau$, to be specific. 
In Fig.~\ref{fig4}~(e) and 
(f) where $\alpha = 4 \times 
10^{-5}$~mm$^{-1}$ ($N = 10^{17}$~cm$^{-3}$), 
$P$ starts to fall appreciably below 0.0028 and 
$\Omega$ begins to modulate and exhibit negative amplitude as the 
optical depth 
increases. As $\alpha$ further increases by a factor of two as 
shown in the subsequent rows in Fig.~\ref{fig4}~(h)and (j), 
$\Omega$ is significantly distorted during the propagation, and the 
laser pulse breaks up into sub-pulses. For further higher value 
of $\alpha$ ($ = 4 \times 
10^{-6}$~mm$^{-1}$), as shown in Fig.~\ref{fig4}~(l), $\Omega$ 
breaks up into sub-pulses with positive and negative amplitudes 
at $k = 100$. Similarly, as shown in Fig.~\ref{fig4}~(g), (i) and 
(k), $P$ saturates at values lower than 0.0028 and exhibits 
Rabi oscillation as optical depth increases. 
We note that, if the pulse area and the pulse duration is much 
shorter than the 
spontaneous decay time which is 2.2~$\mu$s, the propagation 
of the laser pulse follows a non-Beer's law decay profile and 
propagates for much greater optical depths as compared to 
CW laser fields. We expect more distortion if $\alpha$ increases 
further. Clearly, the polarization of the muonic hydrogen 
falls off with increased propagation depth or higher value of $k$, 
with a wiggly behavior due to the modulation of $\Omega$ as 
$k$ increases, as clear from Fig.~\ref{fig4}~(i) and (k) at 
$t = 30$~s. 

\begin{figure*}[!htbp]
\centerline{
\includegraphics[width=\textwidth]{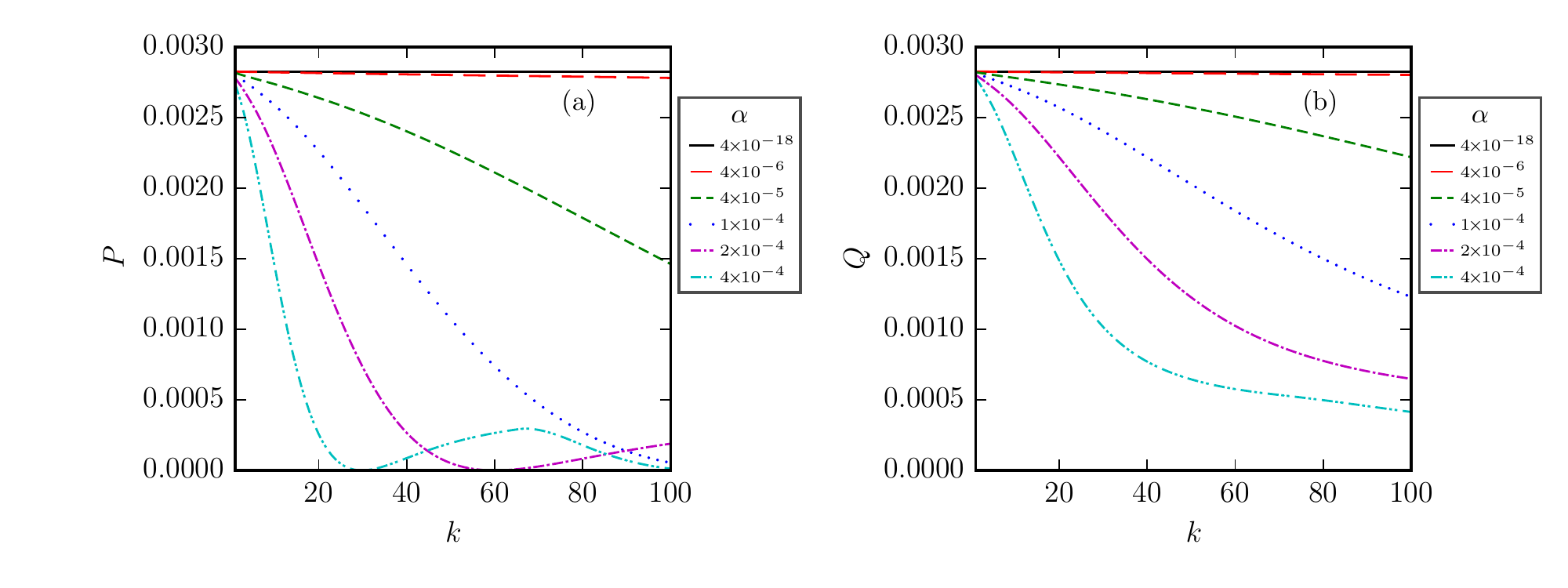}
}
\caption{
(Color online) 
(a) Variation of the population of the ground hyperfine state, $P$, 
(b) variation of the total efficiency, $Q$ 
with the reflection number, $k$ for medium propagation 
coefficient, $\alpha = 4 \times 10^{-18}$ mm$^{-1}$ (black solid line), 
$4 \times 10^{-6}$ mm$^{-1}$ (red long-dashed line), 
$4 \times 10^{-5}$ mm$^{-1}$ (green short-dashed line),
$1 \times 10^{-4}$ mm$^{-1}$ (blue dotted line),
$2 \times 10^{-4}$ mm$^{-1}$ (magenta dot-dashed line), and
$4 \times 10^{-4}$ mm$^{-1}$ (cyan double dot-dashed line). 
}
\label{fig5}
\end{figure*}

To get a clear profile of the polarization of muonic hydrogen atoms 
inside the cavity, in Fig.~\ref{fig5}~(a) and (b), we plot 
the variation of $P$ and the total efficiency, $Q$ as a function 
of $k$, respectively. We define the total efficiency at a 
given value of $k$ 
as the ratio of the total number of excited atoms to the total  
number of atoms in the interaction volume between $k = 0$ to 
the given value of $k$, i.e., 

\begin{align}
\label{eqn:effvsk}
Q & =  \frac{\int_k P_k N a \,dz}{\int_k N a \,dz}.
\end{align}

\noindent
In the above equation, $P_k$ is the value of $P$ at the optical 
depth corresponding to $k$, $N$ is the density of muonic hydrogen 
atoms and $a$ is the beam cross-sectional area after shrinking 
of the beam cross-section. In panel (a)[(b)], we plot 
the values of $P$ ($Q$) for 
$\alpha = 4 \times 10^{-18}$, $4 \times 10^{-6}$, 
$4 \times 10^{-5}$, $1 \times 10^{-4}$, $2 \times 10^{-4}$, and 
$4 \times 10^{-4}$ by black solid, red long-dashed, green 
short-dashed, blue dotted, magenta dot-dashed, and cyan 
double dot-dashed lines, respectively. 
As shown in Fig.~\ref{fig5}~(a), $P$ 
falls off rapidly as 
$\alpha$ increases beyond $4 \times 10^{-6}$ and has a wiggly 
behavior for higher values of $\alpha$ as shown by the 
magenta dot-dashed line and cyan double dot-dashed line in 
Fig.~\ref{fig5}~(a). This is due to the oscillatory behavior of 
$\Omega$ at higher values of $\alpha$. Total 
efficiency, $Q$ also falls off in a similar trend but, the 
wiggly behavior is absent due to the cumulative averaging of $P$ 
over the range of $k$ starting from the entrance of the cavity, 
as shown in Fig.~\ref{fig5}~(b). $Q$ at a certain $k$ gives the 
ratio of the total number of excited atoms to the total number of 
atoms in a target volume accommodating $k$ reflections with the 
beam cross-section shrunk by a factor $k$. Indeed, 
$Q$ as a function 
of $k$ provides information on how the total efficiency varies 
as we shrink the laser beam cross-section and accordingly 
increase the depth of the multipass cavity keeping the target 
volume fixed.

Before closing the discussion, we note that upto a fairly 
high density of the muonic hydrogen atoms for the given pulse 
area of the laser, we can obtain a 
uniformly polarized muonic beam with increased efficiency 
by employing a multipass cavity. 
The so-called threshold density of the muonic hydrogen will 
further rise if we employ a pulse with higher pulse area.
By tailoring the pulse area 
and repetition rate, we can construct the polarized 
muonic beam to be continuous or periodic depending on the 
requirement of the experiment. Keeping the pulse area lower than 
the total optical depth, we can construct a pulsed polarized 
muonic hydrogen beam. Lastly, although the 
the spin-flip probability of the muonic hydrogen atoms 
decreases down the cavity for higher densities of the atoms, 
the total number of atoms excited in the interaction volume is 
always higher for higher density of the atoms. 

\section{Conclusion}

In conclusion, we have studied the prospects of enhancement of 
the laser-induced spin-flip probability between the hyperfine 
ground states of Doppler-broadened muonic hydrogen 
by placing a multipass cavity at a temperature of 10~K. We have 
investigated the effect of propagation of the laser pulse in the 
medium of muonic hydrogen atoms inside the multipass cavity for 
a range of 
densities of the atoms. Assuming a pulse intensity of 
$1 \times 10^6$~W/cm$^2$ and pulse duration of 2~ns, we considered  
the case of shrinking the beam cross-section by a factor of 100 
and employing a multipass cavity which allows 100 reflections of 
the pulse, preserving the total interaction volume.
This transition is dipole-forbidden, and it is the magnetic dipole 
interaction that induces the transition. 
For densities of muonic hydrogen atoms ranging from $10^5$ to 
$\sim 10^{17}$~cm$^{-3}$, the propagation effect is totally 
negligible and the spin-flip probability remains uniform through 
out the cavity and the total efficiency is 0.0028, i.e. 
enhanced by two orders of magnitude. This gives rise to a uniformly 
polarized muon beam. 
As density increases beyond $\sim 10^{17}$~cm$^{-3}$, 
the propagation 
effect can no longer be neglected and spin-flip probability 
is non-uniform inside the cavity. The total efficiency decreases 
as the pulse undergoes more reflections inside the cavity and 
eventually becomes 0.00041 after 100 reflections. This, in turn 
gives rise to a pulsed polarized muonic beam. The result we 
have presented will serve as a guideline for arranging 
muonic beams depending on the requirement of a uniform or pulsed 
polarized muons in $\mu$SR experiments.

\section*{Acknowledgement}
We acknowledge the helpful comments and suggestions of 
Dr. Parvendra Kumar, IIT Chennai, India.

\section*{Competing interests}

The authors declare that there they have no competing interests.

\section*{Funding}

Part of this work was funded by Rashtriya Uchchatar Shiksha 
Abhiyan (RUSA 2.0), MHRD, Govt. of India.

\bigskip

\end{document}